\documentclass[12pt]{iopart}
\usepackage{iopams}  
\usepackage[dvipdfmx]{graphicx}
\usepackage{subfigure}
\usepackage{cite}

\newcommand{\td}{{\rm d}}
\newcommand{\pd}{\partial}
\newcommand{\nab}{\nabla}
\newcommand{\muo}{\mu_{0}}
\newcommand{\hrvz}{\hat{\bi{z}}}
\newcommand{\veps}{\varepsilon}
\newcommand{\vphi}{\varphi}
\newcommand{\vu}{\bi{u}}
\newcommand{\vv}{\bi{v}}
\newcommand{\vx}{\bi{x}}
\newcommand{\vB}{\bi{B}}
\newcommand{\pr}{\prime}
\newcommand{\dis}{\displaystyle}

\begin{document}

\title[Simulated annealing for 
three-dimensional low-beta reduced MHD
]
{
Simulated annealing for  
three-dimensional low-beta\\  reduced MHD equilibria in cylindrical geometry
}

\author{M. Furukawa$^1$ and P. J. Morrison$^2$}

\address{$^1$ Grad. Sch. Eng., Tottori Univ., Minami 4-101, Koyama-cho,
Tottori 680-8552, Japan}
\address{$^2$ Phys. Dept. and Inst. Fusion Studies, Univ. Texas at
Austin, TX, 78712, USA}
\ead{furukawa@damp.tottori-u.ac.jp}
\vspace{10pt}
\begin{indented}
\item[] 26 August 2016
\end{indented}

\begin{abstract}
Simulated annealing (SA) is applied for three-dimensional (3D)
 equilibrium 
 calculation of ideal, low-beta reduced MHD in cylindrical geometry.  
The SA is based on the theory of Hamiltonian mechanics.  The dynamical
 equation of the original system, low-beta reduced MHD in this study, is 
 modified so that the energy changes monotonically while preserving the
 Casimir invariants in the artificial dynamics.  
An equilibrium of the system is given by an
 extremum of the energy, therefore SA can be used as a method for
 calculating ideal MHD equilibrium.  Previous studies demonstrated that
the SA succeeds to lead to various MHD equilibria in two dimensional
 rectangular domain.  In this paper, the theory is applied to
 3D equilibrium of ideal, low-beta reduced MHD.  An
 example of equilibrium with magnetic islands, obtained as a lower
 energy state, is shown.  Several versions of the artificial dynamics
 are developed that can effect smoothing. 
\end{abstract}


\vspace{2pc}

\submitto{\PPCF}

\maketitle
%
%

\section{Introduction}
The calculation of magnetohydrodynamics (MHD) equilibria is fundamental  for fusion plasma research.  Axisymmetric toroidal equilibria are 
 described by the well-known Grad--Shafranov (GS)
equation\cite{Lust-57,Shafranov-58,Grad-58}.  Because the GS equation is an
elliptic differential equation, of  the same type as  Poisson's equation,  numerical methods for solving it are well-established\cite{Lackner-76}. 
The  extension of the GS equation to include plasma rotation has also
received attention\cite{Z-G-72, G-Z-73, Strauss-73,GBMK-04}.  The GS equation including
toroidal rotation is also an elliptic differential equation that  can be 
solved by the same  numerical methods  as the original GS equation.  
Other extensions such as inclusion of anisotropic pressure are
also possible.  (See e.g.\  \cite{Takeda-Tokuda-91} for a review of MHD
equilibrium calculations.)  
The inclusion of poloidal rotation, however, can make the equilibrium
equation hyperbolic\cite{Z-G-72, G-Z-73}, for which no
general numerical method has been established.

The calculation of three-dimensional (3D) MHD equilibria is  considerably  more involved.
The existence of nested magnetic surfaces is not guaranteed generally.
Various numerical codes for the 3D MHD equilibrium have
been developed.   VMEC (Variational Moment Equilibrium Code)\cite{Hirshman-83,
Hirshman-86} may be the most used one, where nested magnetic surfaces
are assumed to exist.  In VMEC the energy of the system is minimized by the
steepest descent method to obtain an equilibrium.  
PIES (Princeton Iterative Equilibrium Solver)\cite{Reiman-86} is another
type, where nested magnetic surfaces are not assumed.  In PIES the solution
method consists of an iteration with the following steps: 
(i) calculation of the pressure by magnetic field line tracing, (ii)
calculation of 
current density by the MHD equilibrium equation
for the obtained pressure and (iii) determination of the magnetic field
by the Amp\`ere's law.   Another code 
HINT (Helical INitial value solver for Toroidal
equilibria)\cite{Harafuji-Hayashi-89} and its spawn  HINT2\cite{Suzuki-06} are 
partly similar to PIES.  HINT2 solves the MHD evolution equation,  
instead of  steps (ii) and (iii) of PIES,  under a  fixed pressure given by  step (i).  
Inclusion of dissipation leads to an equilibrium.  
A new type of the equilibrium code is SPEC (Stepped Pressure Equilibrium
Code)\cite{Hudson-12}, where an equilibrium is constructed by 
connecting multiple layers of Taylor relaxed states (Beltrami
fields) under continuity of the total pressure.  In each layer the
plasma pressure is flat and the existence of magnetic surfaces is not
assumed.   IPEC (Ideal Perturbed Equilibrium Code)\cite{Park-07} 
calculates 3D MHD equilibrium perturbatively by adding
zero-frequency, linear ideal MHD modes to an  axisymmetric equilibrium. 
As for inclusion of plasma rotation, an extension of HINT2 to toroidally
rotating equilibrium is on-going\cite{Suzuki-16}.  

Magnetic island formation, effects of plasma rotation on the magnetic
island, and their interactions with externally applied magnetic
fields  have recently received attention in tokamak as well as
helical\cite{Narushima-15} plasma research.  
Also,  the  transition to helical equilibria  of reversed field pinch
plasmas\cite{Martin-03, Bergerson-11, Piovesan-14} 
is an interesting self-organization phenomenon that is being investigated.
Therefore,  3D MHD equilibrium codes, in addition to nonlinear
evolution codes,  are of great  importance.  For these studies, the existence of magnetic surfaces should not be  assumed, and  plasma rotation should be included.  Moreover, it is
important to characterize  equilibria  in a systematic way
in order to understand important  physical  phenomena.

The present work concerns an MHD equilibrium code of another type based on simulated annealing (SA)\cite{Chikasue-15-PoP, Chikasue-15-JFM}.   
Originally the idea was developed for neutral fluids and demonstrated to work for computing simple equilibria \cite{Vallis-89} (see also \cite{Carnevale-90,Shepherd-90}).  Later it was generalized in \cite{Flierl-Morrison-11} to apply to  a large class of equilibria of Hamiltonian  field theories by allowing for smoothing and the  enforcement of constraints that select out a broader class  equilibria.  SA is based on  Hamiltonian structure, and can be applied to   fluid and plasma models, in particular  ideal magnetohydrodynamics (MHD), because they are  Hamiltonian in terms of  noncanonical Poisson  brackets involving a   skew-symmetric Poisson operator\cite{Morrison-80, Morrison-98}.

For Hamiltonian systems  the time evolution of the dynamical variables is determined by
the functional derivative of the Hamiltonian multiplied by a skew-symmetric Poisson operator.  
The harmonic oscillator is the simplest finite-dimensional example, for which the state variable  is $\vu = (q, p)^{\rm T}$,  with  $q$ and $p$ being the usual canonical coordinate and momentum, respectively, and the 
 Hamiltonian is 
$\dis{
H = ( q^{2} + p^{2} )/2\,.
}$ 
Using   
$\dis{
{\pd H}/{\pd \vu} = ( q, p )^{\rm T}
}$ and 
 skew-symmetric canonical Poisson matrix 
$\dis{
J_c := 
\left(
  \begin{array}{cc}
   0 & 1 \\
   -1 & 0
  \end{array}
\right)\,,
}$ 
gives  the  equations in the Hamiltonian form 
$\dis{
\frac{\td \vu}{\td t} = J_c \frac{\pd H}{\pd \vu}
}$.  
Because of the skew-symmetry of $J_c$, the energy of the oscillator is conserved.  In addition to the energy conservation, for more general noncanonical Poisson brackets, with Poisson operators $J$,  there exist Casimir invariants arising from degeneracy.   Such systems evolve on a surface specified by its energy and the Casimir invariants in the corresponding phase space of the dynamical variables.  The magnetic and cross helicities are  examples for ideal  MHD.   A surface defined by constant Casimir invariants  in the phase space is called
a Casimir leaf.  An extremum of the energy on the Casimir leaf gives an
equilibrium, a stationary state,  as first noted in the plasma literature\cite{KO-58} and then later for the neutral fluid {\cite{Arnold-65-1}.  If we solve the
physical evolution equation, the system follows a trajectory with a
constant energy on the Casimir leaf.  However, it  does not relax to  an equilibrium. 

SA uses  an artificial evolution equation obtained from the Hamiltonian structure of the physical 
 evolution equation by `squaring' the Poisson bracket, i.e. the  dynamics is given by  $\dis{
\frac{\td \vu}{\td t} = J^{2} \frac{\pd H}{\pd \vu}
}$.   For such artificial dynamics, it  is easy to see that the energy
monotonically decreases, as can easily be shown for the harmonic oscillator example. 
Similarly, for MHD, the energy of the system changes monotonically; however, because of the way the artificial evolution equation is based on the Poisson bracket of the physical system, the
Casimir invariants are preserved.  Because  the energy extremum on a Casimir leaf relaxes to an equilibrium  state,  this method can be used as a  numerical method for finding equilibria.  An advantage  of this method is
that the stationary state is characterized by the values of the Casimir invariants.  

The original work \cite{Vallis-89, Carnevale-90,Shepherd-90} was effective for simple equilibria, but because of the plurality of equilibria it did not prove effective for equilibria of interest, and needed to be modified.  This was done in   \cite{Flierl-Morrison-11}, where the term  SA was introduce for this method, by introducing a general  symmetric bracket that  allows for  smoothing and the use of  Dirac theory to  impose  constraints.  On the basis of these early studies,   SA was applied to 2D low-beta reduced MHD\cite{Strauss-76} in \cite{Chikasue-15-PoP}, and a method to pre-adjust values of the Casimir invariants,  by pre-adjusting  initial conditions,  in order to characterize the sought equilibrium states was developed in \cite{Chikasue-15-JFM}.

The previous studies were performed in a 2D rectangular
domain with periodic boundary conditions in both directions, except for
a few cases in \cite{Flierl-Morrison-11} where layer models were used 
for describing a third dimension.  In  the
present study a 3D code is developed, although the outer boundary
of the plasma is still cylindrical.  Then a stationary state with
magnetic islands with multiple helicities can be obtained if it has
lower energy than a cylindrical symmetric state.  The code uses  the
symmetric bracket of \cite{Flierl-Morrison-11} that can effect 
smoothing. 

The paper is organized as follows.  
In section~\ref{sec:theory}, the setting of the problem is explained,  the  SA  method is summarized,  with a focus on the
3D  low-beta reduced MHD example, and  three types of  symmetric brackets are introduced.
 Section~\ref{sec:result} presents numerical results, with the choices of the symmetric brackets  examined, and a  stationary state with magnetic islands  calculated.  Next,  section~\ref{sec:discussion} contains  discussion, where  some  remaining issues are raised.  Finally,  the paper is summarized in
section~\ref{sec:summary}.

\section{Theory}
\label{sec:theory}

\subsection{Reduced MHD system}
\label{subsec:setting}

In this study, let us consider a cylindrical plasma with  minor
radius $a$ and length $2 \pi R_{0}$.  The cylindrical coordinates are
$(r, \theta, z)$, with the  toroidal angle being 
$\dis{
\zeta := {z}/{R_{0}}
}$ and the inverse aspect ratio  given by 
$\dis{
\veps := {a}/{R_{0}}
}$.  
Physical quantities are normalized by the length $a$, the magnetic field
in the $z$-direction $B_{0}$, the Alfv\'en velocity 
$\dis{
v_{\rm A} :=  {B_{0}}/{\sqrt{ \muo \rho_{0} }}
}$ 
with $\muo$ and $\rho_{0}$ being vacuum permeability and typical mass
density, respectively, and the Alfv\'en time 
$\dis{
\tau_{\rm A} := {a}/{v_{\rm A}}
}$.  
Then  low-beta reduced MHD is given by 
\begin{eqnarray}
 \frac{\pd U}{\pd t}
=
 [ U , \vphi ] + [ \psi , J ] - \veps \frac{\pd U}{\pd \zeta},
\label{eq:vorticity-equation}
\\
 \frac{\pd \psi}{\pd t}
=
 [ \psi , \vphi ] - \veps \frac{\pd \vphi}{\pd \zeta},
\label{eq:Ohm-law}
\end{eqnarray}
where 
the fluid velocity is $\vv = \hrvz \times \nab \vphi$, 
the magnetic field is $\vB = \hrvz + \nab \psi \times \hrvz$, 
the vorticity is $U := \bigtriangleup_{\perp} \vphi$, 
the current density is $J := \bigtriangleup_{\perp} \psi$,
the Poisson bracket for two functions $f$ and $g$ is $[ f, g ] := \hrvz
\cdot \nab f \times \nab g$, 
the unit vector in the $z$ direction
is denoted by $\hrvz$, 
and $\bigtriangleup_{\perp}$ is the Laplacian in the $r$--$\theta$
plane.  

\subsection{Simulated annealing theory}
\label{subsec:theory-SA}

Now,  we briefly  review  the governing  SA system, referring the  reader to 
 \cite{Flierl-Morrison-11} for a detailed explanation.
The artificial dynamics of SA  is given by
\begin{equation}
 \frac{\pd \vu}{\pd t}
=
 (( \vu, H )),
\label{eq:evolution-equation-SA}
\end{equation}
where $\vu$ is a vector of the dynamical variables, 
$H[\vu]$ is the Hamiltonian functional 
and $(( F, G ))$ is the symmetric bracket for two functionals 
$F[\vu]$ and $G[\vu]$,  defined by
\begin{equation}
 (( F, G ))
:=
 \int_{\cal D} \td^{3}x^{\pr}
 \int_{\cal D} \td^{3}x^{\pr\pr} \,
 \{ F, u^{i}(\vx^{\pr}) \}
 K_{ij}(\vx^{\pr}, \vx^{\pr\pr})
 \{ u^{j}(\vx^{\pr\pr}), G \},
\label{eq:symmetric-bracket}
\end{equation}
where $( K_{ij} )$ is a definite symmetric kernel, and 
\begin{equation}
 \{ F, G \}
:= 
 \int_{\cal D} \td^{3}x^{\pr}
 \int_{\cal D} \td^{3}x^{\pr\pr} \,
 \frac{\delta F[\vu]}{\delta u^{i}(\vx^{\pr})}
 J^{ij}( \vx^{\pr}, \vx^{\pr\pr} )
 \frac{\delta G[\vu]}{\delta u^{j}(\vx^{\pr\pr})}
\label{eq:Poisson-bracket}
\end{equation}
is the Poisson bracket for two functionals, with  ${\cal D}$ denoting  the  
whole domain of the system.  The quantity 
$( J^{ij} )$ is the skew-symmetric Poisson operator, and  
$\dis{
{\delta F[\vu]}/{\delta \vu}
}$ 
and 
$\dis{
{\delta G[\vu]}/{\delta \vu}
}$ 
are the functional derivatives of $F$ and $G$, respectively.   The sign of the right-hand side is taken so  that energy decreases   as time progresses.  

The Hamiltonian structure for low-beta reduced MHD, as  was first given in \cite{MH-84,MM-84}, has $\vu := ( u^{1}, u^{2} )^{\rm T}$ where $u^{1} = U$ and $u^{2} = \psi$, and the Hamiltonian 
\begin{equation}
 H[ \vu ] 
:= 
 \int_{\cal D} \td^{3}x \,
      \frac{1}{2} \left\{
         \left| \nab_{\perp} ( \bigtriangleup_{\perp}^{-1} U ) \right|^{2}
       + \left| \nab \psi \right|^{2}
                  \right\},
\label{eq:Hamiltonian}
\end{equation}
with  ${\cal D}$ being  the whole domain of the cylindrical plasma.  
The first and the second terms of the  the integrand of (\ref{eq:Hamiltonian}) are  the kinetic energy
$E_{\rm k}$ and magnetic energy $E_{\rm m}$, respectively, and the  skew-symmetric Poisson operator is given by
\begin{equation}
\fl
 ( J^{ij}(\vx^{\pr}, \vx^{\pr\pr}) )
=
 \delta^{3}(\vx^{\pr} - \vx^{\pr\pr})
 \left(
   \begin{array}{cc}
    -[ U(\vx^{\pr\pr}), \quad ]
    & 
    -[ \psi(\vx^{\pr\pr}), \quad ] + \veps \frac{\pd }{\pd \zeta^{\pr\pr}}
    \\
    -[ \psi(\vx^{\pr\pr}), \quad ] + \veps \frac{\pd }{\pd \zeta^{\pr\pr}}
    & 
    0
   \end{array}
 \right).
\label{eq:J-operator}
\end{equation}

In order to write down the evolution equation of the SA, we need to
calculate the Poisson bracket between the dynamical variables, and
between the dynamical variable and the Hamiltonian.  
The functional derivative of $H[\vu]$ is straightforward: 
\begin{equation}
 \frac{\delta H[\vu]}{\delta \vu}
=
 \left(
  \begin{array}{c}
   -\vphi
   \\
   -J
  \end{array}
 \right).
\label{eq:functional-derivative-H}
\end{equation}
The functional derivatives of $U$ and $\psi$ can be obtained by
considering the  functionals 
\begin{eqnarray}
 U(\vx)
= 
 \int_{\cal D} \td^{3}x^{\pr} \, U(\vx^{\pr}) \delta^{3}(\vx - \vx^{\pr}),
\label{eq:U-functional}
\\
 \psi(\vx)
= 
 \int_{\cal D} \td^{3}x^{\pr} \, \psi(\vx^{\pr}) \delta^{3}(\vx - \vx^{\pr}),
\label{eq:psi-functional}
\end{eqnarray}
where $\delta^{3}(\vx)$ is the Dirac's delta function in
three-dimensional space, 
and its explicit form is 
$\dis{
\delta^{3}(\vx) =  \delta(r) \delta(\theta) \delta(\zeta)\, {\veps}/{r}
}$ 
since 
$\dis{
\td^{3}x = \td r \td \theta \td \zeta \, {r}/{\veps}
}$.  
The functional derivatives of $U(\vx)$ and $\psi(\vx)$ are
\begin{eqnarray}
 \frac{\delta U(\vx)}{\delta \vu}
=
 \left(
  \begin{array}{c}
   \delta^{3}(\vx - \vx^{\pr})
   \\
   0
  \end{array}
 \right),
\label{eq:functional-derivative-U}
\\
 \frac{\delta \psi(\vx)}{\delta \vu}
=
 \left(
  \begin{array}{c}
   0
   \\
   \delta^{3}(\vx - \vx^{\pr})
  \end{array}
 \right)\,, 
\label{eq:functional-derivative-psi}
\end{eqnarray}
whence we obtain
\begin{eqnarray}
 \{ U(\vx), U(\vx^{\pr}) \}
=
 [ U(\vx^{\pr}), \delta^{3}(\vx - \vx^{\pr}) ],
\label{eq:Poisson-bracket-U-U}
\\
 \{ U(\vx), \psi(\vx^{\pr}) \}
=
 [ \psi(\vx^{\pr}), \delta^{3}(\vx - \vx^{\pr}) ] 
 - \veps \frac{\pd \delta^{3}(\vx - \vx^{\pr})}{\pd \zeta^{\pr}},
\label{eq:Poisson-bracket-U-psi}
\\
 \{ \psi(\vx), U(\vx^{\pr}) \}
=
 [ \psi(\vx^{\pr}), \delta^{3}(\vx - \vx^{\pr}) ] 
 - \veps \frac{\pd \delta^{3}(\vx - \vx^{\pr})}{\pd \zeta^{\pr}},
\label{eq:Poisson-bracket-psi-U}
\\
 \{ \psi(\vx), \psi(\vx^{\pr}) \}
= 
 0,
\label{eq:Poisson-bracket-psi-psi}
\\
 \{ U(\vx^{\pr\pr}), H \}
=
 [ U(\vx^{\pr\pr}), \vphi(\vx^{\pr\pr}) ] 
 + [ \psi(\vx^{\pr\pr}), J(\vx^{\pr\pr}) ] 
 - \veps \frac{\pd J(\vx^{\pr\pr})}{\pd \zeta^{\pr\pr}},
\label{eq:Poisson-bracket-U-H}
\\
 \{ \psi(\vx^{\pr\pr}), H \}
=
 [ \psi(\vx^{\pr\pr}), \vphi(\vx^{\pr\pr}) ] 
 - \veps \frac{\pd \vphi(\vx^{\pr\pr})}{\pd \zeta^{\pr\pr}}.
\label{eq:Poisson-bracket-psi-H}
\end{eqnarray}
These confirm that the physical evolution equations are written as
\begin{equation}
 \frac{\pd \vu}{\pd t} 
=
 \{ \vu, H \}.
\label{eq:evolution-equation-physical}
\end{equation}

Using the above  Poisson brackets, the symmetric brackets are obtained as
\begin{eqnarray}
 (( U, H ))
=
 [ U(\vx), \tilde{\vphi}(\vx) ]
 + [ \psi(\vx), \tilde{J}(\vx) ]
 - \veps \frac{\pd \tilde{J}(\vx)}{\pd \zeta},
\label{eq:symmetric-bracket-U-H}
\\
 (( \psi, H ))
=
 [ \psi(\vx), \tilde{\vphi}(\vx) ]
 - \veps \frac{\pd \tilde{\vphi}(\vx)}{\pd \zeta},
\label{eq:symmetric-bracket-psi-H}
\end{eqnarray}
where
\begin{eqnarray}
 \tilde{\vphi}(\vx)
=
 \int_{\cal D} \td^{3}x^{\pr\pr} \,
   \left(
     K_{UU}(\vx, \vx^{\pr\pr}) f^{U}(\vx^{\pr\pr})
     + K_{U \psi}(\vx, \vx^{\pr\pr}) f^{\psi}(\vx^{\pr\pr})
   \right),
\label{eq:tvphi-general}
\\
 \tilde{J}(\vx)
=
 \int_{\cal D} \td^{3}x^{\pr\pr} \,
   \left(
     K_{\psi U}(\vx, \vx^{\pr\pr}) f^{U}(\vx^{\pr\pr})
     + K_{\psi \psi}(\vx, \vx^{\pr\pr}) f^{\psi}(\vx^{\pr\pr})
   \right)\,, 
\label{eq:tJ-general}
\end{eqnarray}
and
\begin{equation}
 ( K_{ij}(\vx^{\pr}, \vx^{\pr\pr}) )
=
 \left(
  \begin{array}{cc}
   K_{UU}(\vx^{\pr}, \vx^{\pr\pr})
   & 
   K_{U\psi}(\vx^{\pr}, \vx^{\pr\pr})
   \\
   K_{\psi U}(\vx^{\pr}, \vx^{\pr\pr})
   & 
   K_{\psi \psi}(\vx^{\pr}, \vx^{\pr\pr})
  \end{array}
 \right), 
\label{eq:symmetric-kernel-general}
\end{equation}
with $f^{U}$ and $f^{\psi}$ being defined by the right-hand sides of the 
physical evolution equation multiplied by the negative sign as
\begin{eqnarray}
 f^{U}(\vx)
:=
 - \left(
   [ U(\vx), \vphi(\vx) ] 
 + [ \psi(\vx), J(\vx) ]
 - \veps \frac{\pd J(\vx)}{\pd \zeta}
   \right),
\label{eq:fU}
\\
 f^{\psi}(\vx)
:=
 -\left(
 [ \psi(\vx), \vphi(\vx) ]
 - \veps \frac{\pd \vphi(\vx)}{\pd \zeta}
  \right).
\label{eq:fpsi}
\end{eqnarray}

Observe, $U$ and $\psi$ are advected by $\tilde{\vphi}$ and
$\tilde{J}$ in  SA, instead of $\vphi$ and $J$ in the physical
dynamics. 
A variety of  artificial dynamics can be generated by different choices  of the kernel $( K_{ij} )$.

\subsection{Casimir invariants}
\label{subsec:Casimir}

Before examining the choice of $( K_{ij} )$, let us introduce the
Casimir invariants.  
A Casimir invariant of the system is defined as a functional
$C[\vu]$ that satisfies
\begin{equation}
 \{ C, F \} = 0
\label{eq:Casimir-definition}
\end{equation}
for any functional $F[\vu]$.  If we write 
$\dis{
{\delta C[\vu]}/{\delta \vu} = ( C_{1}, C_{2} )^{\rm T}
}$ 
and 
$\dis{
{\delta F[\vu]}/{\delta \vu} = ( F_{1}, F_{2} )^{\rm T}
}$,
then 
\begin{eqnarray}
\hspace*{-1.5cm}
 \{ C, F \}
&=
 \int_{\cal D} \td^{3}x^{\pr}
 \int_{\cal D} \td^{3}x^{\pr\pr} \,
 \delta^{3}(\vx^{\pr} - \vx^{\pr\pr}) 
\nonumber
\\
& \qquad 
 \left(
 C_{1}(\vx^{\pr}) \left(
   -[ U(\vx^{\pr\pr}), F_{1}(\vx^{\pr\pr}) ]
   -[ \psi(\vx^{\pr\pr}), F_{2}(\vx^{\pr\pr}) ]
   + \veps \frac{\pd F_{2}(\vx^{\pr\pr})}{\pd \zeta^{\pr\pr}}
                  \right)
 \right.
\nonumber
\\
& \qquad 
 \left.
 + C_{2}(\vx^{\pr}) \left(
   -[ \psi(\vx^{\pr\pr}), F_{1}(\vx^{\pr\pr}) ]
   + \veps \frac{\pd F_{1}(\vx^{\pr\pr})}{\pd \zeta^{\pr\pr}}
                  \right)
                             \right)
\nonumber
\\
&=
 \int_{\cal D} \td^{3}x^{\pr} \,
\left(
 C_{1}(\vx^{\pr}) \left(
   -[ U(\vx^{\pr}), F_{1}(\vx^{\pr}) ]
   -[ \psi(\vx^{\pr}), F_{2}(\vx^{\pr}) ]
   + \veps \frac{\pd F_{2}(\vx^{\pr})}{\pd \zeta^{\pr}}
                  \right)
\right.
\nonumber
\\
& \qquad \qquad \qquad 
\left.
 + C_{2}(\vx^{\pr}) \left(
   -[ \psi(\vx^{\pr}), F_{1}(\vx^{\pr}) ]
   + \veps \frac{\pd F_{1}(\vx^{\pr})}{\pd \zeta^{\pr}}
                    \right)
\right)
\nonumber
\\
&=
 \int_{\cal D} \td^{3}x^{\pr} \,
\left(
 F_{1}(\vx^{\pr}) \left(
   -[ C_{1}(\vx^{\pr}), U(\vx^{\pr}) ]
   -[ C_{2}(\vx^{\pr}), \psi(\vx^{\pr}) ]
   - \veps \frac{\pd C_{2}(\vx^{\pr})}{\pd \zeta^{\pr}}
                  \right)
\right.
\nonumber
\\
& \qquad \qquad \qquad
\left.
 + F_{2}(\vx^{\pr}) \left(
   -[ C_{1}(\vx^{\pr}), \psi(\vx^{\pr}) ]
   - \veps \frac{\pd C_{1}(\vx^{\pr})}{\pd \zeta^{\pr}}
                  \right)
\right)\,, 
\end{eqnarray}
where the last equality follows upon integration by parts.
In order to satisfy $\{ C, F \} = 0$ for any $F_{1}$ and $F_{2}$, 
$C_{1}$ and $C_{2}$ must satisfy
\begin{eqnarray}
  && [ C_{1}(\vx), U(\vx) ]
   +[ C_{2}(\vx), \psi(\vx) ]
   + \veps \frac{\pd C_{2}(\vx)}{\pd \zeta}
= 0,
\label{eq:Casimir-condition-1}
 \\
 &&  [ C_{1}(\vx), \psi(\vx) ]
   + \veps \frac{\pd C_{1}(\vx)}{\pd \zeta}
= 0.
\label{eq:Casimir-condition-2}
\end{eqnarray}
Choosing  $C_{1} = 0$ and $C_{2} = 1$,  yields 
$\dis{
C = \int_{\cal D} \td^{3}x \, \psi(\vx) =: C_{\rm m}
}$,  while choosing 
$C_{1} = 1$ and $C_{2} = 0$, 
yields 
$\dis{
C = \int_{\cal D} \td^{3}x \, U(\vx) =: C_{\rm v}
}$. (See  \cite{MH-84} for a discussion of how these are remnants of the helicity and cross helicity.) 
The accuracy of a numerical simulation can be tested by monitoring the conservation of
$C_{\rm m}$ and $C_{\rm v}$.

\subsection{Cross helicity}
\label{subsec:cross-helicity}

Another conserved quantity of ideal MHD is a cross helicity, which is
defined by
\begin{eqnarray}
 C_{\rm c} 
&:=&
 \int_{\cal D} \td^{3}x \, \vv \cdot \vB
\\
&=&
 \int_{\cal D} \td^{3}x \, U \psi.
\label{eq:cross-helicity}
\end{eqnarray}
The functional derivative of $C_{\rm c}$ is given by
\begin{equation}
 \frac{\delta C_{\rm c}}{\delta \vu}
=
 \left(
  \begin{array}{c}
   \psi
   \\
   U
  \end{array}
 \right).
\label{eq:functional-derivative-cross-helicity}
\end{equation}
Then, $[C_{1}(\vx), U(\vx)] + [C_{2}(\vx), \psi(\vx)] = 0$ and 
$[C_{1}(\vx), \psi(\vx)] = 0$ in (\ref{eq:Casimir-condition-1}) and
(\ref{eq:Casimir-condition-2}), however,  
$\zeta$-derivative terms remains finite generally.  
When $F[\vu]$ is taken to be $H[\vu]$, $F_{1} = -\vphi$ and $F_{2} =
-J$, and we can show that
\begin{eqnarray}
 \{ C_{\rm c}, H \}
&=
 \int_{\cal D} \td^{3}x^{\pr} \,
 \left(
    \vphi(\vx^{\pr}) \veps \frac{\pd U(\vx^{\pr})}{\pd \zeta^{\pr}}
  + J(\vx^{\pr}) \veps \frac{\pd \psi(\vx^{\pr})}{\pd \zeta^{\pr}}
 \right)
\nonumber
\\
&=
 -\frac{1}{2} 
 \int_{\cal D} \td^{3}x^{\pr} \,
 \frac{\pd}{\pd \zeta^{\pr}}
 \left(
   | \nab_{\perp} \vphi |^{2}
   + | \nab_{\perp} \psi |^{2}
 \right)
\nonumber
\\
&= 0
\end{eqnarray}
by integration by parts under appropriate boundary conditions.
Therefore, the cross helicity $C_{\rm c}$ is also conserved by the SA as
well as the physical dynamics.

\subsection{Choices for the symmetric kernel}
\label{subsec:kernel}

In subsection \ref{subsec:theory-SA}, we obtained a general form of
$\tilde{\vphi}$ and $\tilde{J}$ in (\ref{eq:tvphi-general}) and
(\ref{eq:tJ-general}).  Here we  introduce three choices for 
$( K_{ij} )$.  
First, let us set the off-diagonal terms of $( K_{ij} )$ to zero in this
paper.    Then, we may use  the definition  
\begin{equation}
 \tilde{h}(\vx)
=
 \int_{\cal D} \td^{3}x^{\pr\pr} \, 
  K(\vx, \vx^{\pr\pr}) f(\vx^{\pr\pr}),
\label{eq:th-general}
\end{equation}
with  $\tilde{h}$ chosen to be $\tilde{\vphi}$ or $\tilde{J}$, $K$ to be 
 $K_{UU}$ or $K_{\psi \psi}$, and $f$    to  be 
$f^{U}$ or $f^{\psi}$, respectively.

For our  \underline{first} choice of smoothing we consider 
\begin{equation}
 ( K_{ij}(\vx, \vx^{\pr\pr}) )
=
 \left(
  \begin{array}{cc}
   \alpha_{UU}\,  \delta^{3}(\vx - \vx^{\pr\pr})
   & 
   0
   \\
   0
   & 
   \alpha_{\psi \psi} \, \delta^{3}(\vx - \vx^{\pr\pr})
  \end{array}
 \right),
\label{eq:symmetric-kernel-SA1}
\end{equation}
where $\alpha_{UU} > 0$ and $\alpha_{\psi \psi} > 0$ are constants that 
scale the resulting  advection fields $\tilde{\vphi}$ and $\tilde{J}$.
Then it is straightforward to obtain $\tilde{h}(\vx) = \alpha f(\vx)$;
namely, 
\begin{eqnarray}
 \tilde{\vphi}(\vx)
=
 \alpha_{UU} f^{U}(\vx),
\label{eq:tvphi-SA1}
\\
 \tilde{J}(\vx)
=
 \alpha_{\psi \psi} f^{\psi}(\vx),
\label{eq:tJ-SA1}
\end{eqnarray}
where $\alpha$ represents $\alpha_{UU}$ or $\alpha_{\psi\psi}$,
respectively, and similarly for  $\tilde{h}$ etc.  
These advection fields are the right-hand sides of the physical
evolution equations (\ref{eq:vorticity-equation}) and (\ref{eq:Ohm-law})
multiplied by $-\alpha_{UU}$ and $-\alpha_{\psi \psi}$, respectively.  
We refer to  this version of smoothing as  ``SA-1''. 

The \underline{second} choice of smoothing  introduced in this paper is
\begin{equation}
\fl
 ( K_{ij}(\vx, \vx^{\pr\pr}) )
=
 \left(
  \begin{array}{cc}
   \alpha_{UU} \, {\veps} \, \delta(r - r^{\pr\pr}) 
   g_{\theta\zeta}(\theta, \zeta, \theta^{\pr\pr}, \zeta^{\pr\pr}) /r
   & 
   0
   \\
   0
   & 
   \alpha_{\psi \psi} \, {\veps}\,  \delta(r - r^{\pr\pr}) 
   g_{\theta\zeta}(\theta, \zeta, \theta^{\pr\pr}, \zeta^{\pr\pr})/r
  \end{array}
 \right),
\label{eq:symmetric-kernel-SA2}
\end{equation}
where $\delta(r - r^{\pr\pr})$ is a Dirac's delta function in $r$,
and $g_{\theta \zeta}$ is defined by 
\begin{equation}
 \left( 
    \frac{\pd^{2}}{\pd \theta^{2}} 
   +\frac{\pd^{2}}{\pd \zeta^{2}} 
  \right)
  g_{\theta\zeta}(\theta, \zeta, \theta^{\pr\pr}, \zeta^{\pr\pr})
= - 
  \delta(\theta - \theta^{\pr\pr})
  \delta(\zeta - \zeta^{\pr\pr}),
\label{eq:Poisson-eq-2D}
\end{equation}
i.e., $g_{\theta\zeta}$ is a Green's function in the $\theta$--$\zeta$ plane.
Now, if we Fourier expand $\tilde{h}(\vx)$ in $\theta$ and $\zeta$ as
\begin{equation}
 \tilde{h}(\vx) 
= 
 \sum_{m,n} \tilde{h}_{m/n}(r) \rme^{\rmi ( m \theta + n \zeta)}, 
\end{equation}
then the Fourier coefficients are given by
\begin{eqnarray}
 \tilde{h}_{m/n}(r) 
&= 
 \frac{1}{(2 \pi)^{2}}
 \oint \td \theta \oint \td \zeta \,
 \tilde{h}(\vx) \rme^{-\rmi ( m \theta + n \zeta )}
\nonumber
\\
&=
 \frac{1}{(2 \pi)^{2}}
 \oint \td \theta \oint \td \zeta \,
 \int_{\cal D} \td^{3}x^{\pr\pr} \, 
  K(\vx, \vx^{\pr\pr}) f(\vx^{\pr\pr})
 \rme^{-\rmi ( m \theta + n \zeta )}
\nonumber
\\
&=
 \int_{\cal D} \td^{3}x^{\pr\pr} \, 
 f(\vx^{\pr\pr}) 
 \frac{1}{(2 \pi)^{2}}
 \oint \td \theta \oint \td \zeta \,
 K(\vx, \vx^{\pr\pr}) \rme^{-\rmi ( m \theta + n \zeta )}
\nonumber
\\
&=
 \int_{\cal D} \td^{3}x^{\pr\pr} \, 
 f(\vx^{\pr\pr}) K_{m/n}(r, \vx^{\pr\pr}).
\label{eq:thmn-general}
\end{eqnarray}
By Fourier transforming (\ref{eq:Poisson-eq-2D}) in $\theta$ and $\zeta$
to obtain the Fourier expansion coefficients of $g_{\theta\zeta}$, and
we obtain
\begin{equation}
 K_{m/n}(r, \vx^{\pr\pr})
=
 \delta(r - r^{\pr\pr})
 \frac{ \alpha\, \veps}{(2 \pi)^{2} (m^{2} + n^{2})\,  r} \, 
 \rme^{-\rmi ( m \theta^{\pr\pr} + n \zeta^{\pr\pr} )}.
\label{eq:Kmn-SA2}
\end{equation}
Then 
\begin{equation}
 \tilde{h}_{m/n}(r)
=
\frac{ \alpha \, \veps}{m^{2} + n^{2}}\,  f_{m/n}(r)\,, 
\label{eq:hmn-SA2}
\end{equation}
which  gives the advection fields $\tilde{\vphi}$ and
$\tilde{J}$.
The symmetric bracket  with this smoothing has the effect of  reducing  short-wave-length
components of the advection fields in the $\theta$--$\zeta$ plane,
and  is similar to the one introduced for 2D vortex
dynamics\cite{Flierl-Morrison-11} and 2D low-beta reduced
MHD\cite{Chikasue-15-PoP, Chikasue-15-JFM}.  
We refer to  this version of smoothing as  ``SA-2''. . 

Lastly, consider our \underline{third} choice for smoothing, 
\begin{equation}
 ( K_{ij}(\vx, \vx^{\pr\pr}) )
=
 \left(
  \begin{array}{cc}
   \alpha_{UU}\,  g(\vx, \vx^{\pr\pr})
   & 
   0
   \\
   0
   & 
   \alpha_{\psi \psi}\,  g(\vx, \vx^{\pr\pr})
  \end{array}
 \right),
\label{eq:symmetric-kernel-SA3}
\end{equation}
where 
\begin{equation}
 \bigtriangleup g(\vx, \vx^{\pr\pr})
:= -
  \delta^{3}(\vx - \vx^{\pr\pr}),
\label{eq:Poisson-eq-3D}
\end{equation}
i.e., each diagonal component of the symmetric kernel is proportional to the  Green's
function in 3D.  
If we Fourier expand $g$ in $\theta$, $\zeta$, $\theta^{\pr\pr}$ and
$\zeta^{\pr\pr}$ as
\begin{equation}
 g(\vx, \vx^{\pr\pr})
=
 \sum_{m,n} \sum_{m^{\pr\pr}, n^{\pr\pr}}
 g_{m/n, m^{\pr\pr}/n^{\pr\pr}} (r, r^{\pr\pr})
 \rme^{\rmi ( m \theta + n \zeta )}
 \rme^{\rmi ( m^{\pr\pr} \theta^{\pr\pr} + n^{\pr\pr} \zeta^{\pr\pr} )}, 
\end{equation}
we can express (\ref{eq:Poisson-eq-3D})
as
\begin{equation}
\fl
 \frac{1}{r} \frac{\pd}{\pd r} 
 \left(
   r \frac{\pd}{\pd r} \left( g_{m/n,-m/-n}(r, r^{\pr\pr}) \right)
 \right)
+ \left(
    \frac{m^{2}}{r^{2}} + \veps^{2} n^{2}
   \right)
  g_{m/n,-m/-n}(r, r^{\pr\pr})
=
 \frac{-\veps}{(2 \pi)^{2}r}  
 \delta(r - r^{\pr\pr}).
\label{eq:Poisson-eq-3D-mn}
\end{equation}
Here $g_{m/n, -m/-n}(r, r^{\pr\pr})$ means 
$g_{m/n, m^{\pr\pr}/n^{\pr\pr}}(r, r^{\pr\pr})$ with $m^{\pr\pr}=-m$ and 
$n^{\pr\pr}=-n$.
For a given $r^{\pr\pr}$, we can solve the homogeneous equation to
obtain the solutions in both $0 \leq r < r^{\pr\pr}$ and $r^{\pr\pr} < r
\leq 1$ regions.  These are actually linear combinations of the modified
Bessel functions.  In order to determine the coefficients of the linear
combination, we require the continuity of $g_{m/n, -m/-n}$ and the jump condition
\begin{equation}
 r \left. \frac{\pd g_{m/n,-m/-n}(r, r^{\pr\pr})}{\pd r} 
    \right\vert_{r^{\pr\pr}-0}^{r^{\pr\pr}+0}
= -
 \veps
\label{eq:g-jump-condition}
\end{equation}
at $r = r^{\pr\pr}$.
The jump condition (\ref{eq:g-jump-condition}) is obtained by integrating
(\ref{eq:Poisson-eq-3D-mn}) from $r^{\pr\pr} - 0$ to $r^{\pr\pr}+0$.
By using this symmetric kernel, we obtain
\begin{equation}
 \tilde{h}_{m/n}(r)
=
 \alpha
 \frac{(2 \pi)^{2}}{\veps}
 \int_{0}^{1} \td r^{\pr\pr} \,
 g_{m/n,-m/-n}(r, r^{\pr\pr}) f_{m/n}(r^{\pr\pr}).
\label{eq:hmn-SA3}
\end{equation}
This version of  smoothing can effect not only the behavior in the  $\theta$--$\zeta$ plane
but also in the $r$ direction.
We refer to  this version of smoothing as  ``SA-3''.

\section{Numerical results}
\label{sec:result}

Consider now our  numerical results obtined from a code developed for solving the artificial evolution
equation (\ref{eq:evolution-equation-SA}).  The code imposes regularity of  
physical quantities at $r=0$ and $\vphi = \psi = 0$  at the plasma boundary. 
The pseudo-spectral method is adopted in $\theta$ and $\zeta$, which allows for  multiple helicities,  while a  second-order finite difference method is used in $r$.    For  time advancement, fourth-order Runge--Kutta  with
step-size control is used.   Starting from an initial condition, the artificial
evolution equation is solved and,  in accordance with theory,  the energy of the system decreases
monotonically.  When the relative change rates of both kinetic and
magnetic energy, 
$\dis{
{| \td E_{\rm k} / \td t |}/{E_{\rm k}}
}$
and 
$\dis{
{| \td E_{\rm m} / \td t |}/{E_{\rm m}}
}$,
become smaller than a tolerance, the simulation is stopped.  

For the numerical results shown below, 
the inverse aspect ratio $\veps = 1/10$, while the  grid numbers for  $r$, $\theta$
and $\zeta$  are $100$, $32$ and $16$, respectively.  The
Fourier mode numbers included in the simulation are $-10 \leq m \leq 10$
and $0 \leq n \leq 5$, respectively.  
The tolerance for the convergence was chosen to be $10^{-6}$.

We present results for  two initial conditions.  
The first corresponds to a trivial stationary state where 
$U = U(r)$ and $\psi = \psi(r)$, with corresponding  $\vphi$ and $J$ satisfying  $U = \bigtriangleup_{\perp} \vphi$ and  $J = \bigtriangleup_{\perp} \psi$,  also being 
functions of $r$ only.  Clearly, the right-hand side of
(\ref{eq:evolution-equation-SA}), or
(\ref{eq:evolution-equation-physical}), becomes zero in this case, and
no change of the system occurs.   Indeed, the simulation code stopped immediately after initializing. 

The second initial condition  is given by the stationary state of the first one,  plus a
small perturbation that has a radial magnetic field resonant at a rational 
surface.  The small perturbation changes the field-line topology by opening a magnetic island.
If the stationary state with cylindrical symmetry is
unstable against the associated  tearing mode, we expect  the
system to  evolve and reach a stationary state with magnetic islands,
with  its energy  decreased by the SA.  

Figure~\ref{fig:equilibrium} shows the safety factor profile $q(r)$  of
the stationary state with cylindrical symmetry.  The plasma rotation was
assumed to be zero and   $\psi(r)$ was chosen so  that the safety factor 
$\dis{
q(r) = - {\veps r}/{\psi^{\pr}(r)}
}$, 
where the prime denotes $r$ derivative.
Specifically, 
$
\dis{
q(r)
= 
{q_{0}}/({1 - {r^{2}}/{2}})
}$ 
was used, where $q_{0}$ is the safety factor at $r=0$, which gives 
 $q=2$ at 
$\dis{
r={1}/{2}
}$.

\begin{figure}[h]
 \centering
 \includegraphics[width=0.47\textwidth]{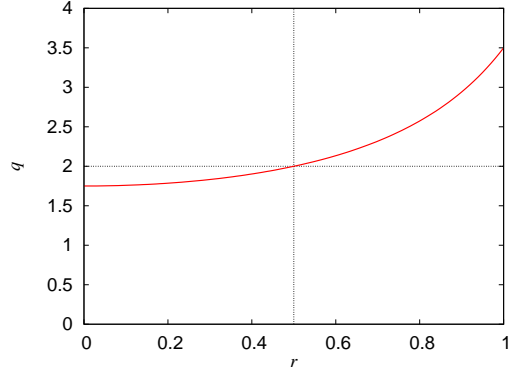}
 \caption{The safety factor profile $q(r)$ for  a stationary state.}
 \label{fig:equilibrium}
\end{figure}

The stationary state shown in figure~\ref{fig:equilibrium} is unstable
against a  tearing mode with  mode numbers  $m=-2$ and $n=1$, which has   
the tearing mode parameter\cite{F-K-R-63}  $\Delta^{\pr} \approx  22.4$.  
Thus  a small perturbation with $m=-2$ and $n=1$ was added to the
cylindrically symmetric state, giving a  radial magnetic
field across the $q=2$ surface. 
The radial profiles of the $m=-2$ and $n=1$ components are shown in 
figure~\ref{fig:initial-perturbation}.

\begin{figure}[h]
 \centering
 \subfigure[]{
   \includegraphics[width=0.47\textwidth]{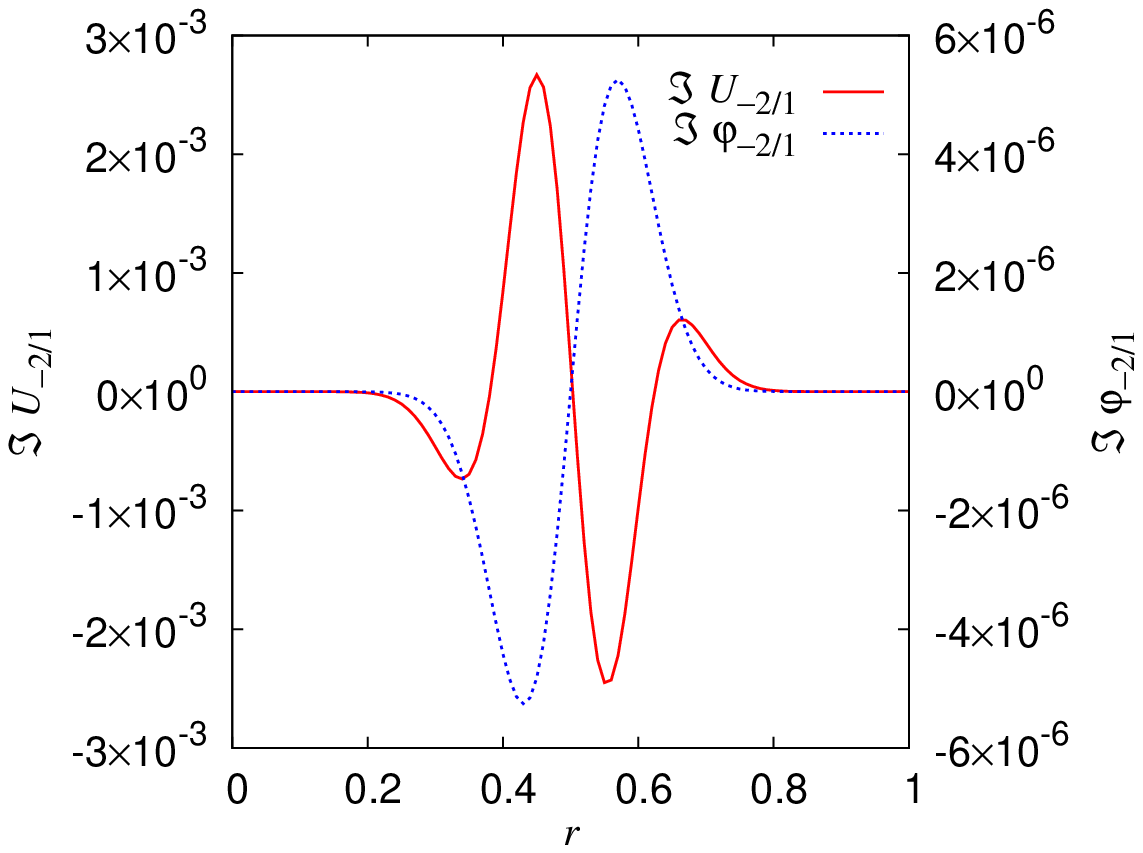}
   \label{fig:r-U_i-phi_i-mm2-n1-t000000_0000}
 } 
 \subfigure[]{
   \includegraphics[width=0.47\textwidth]{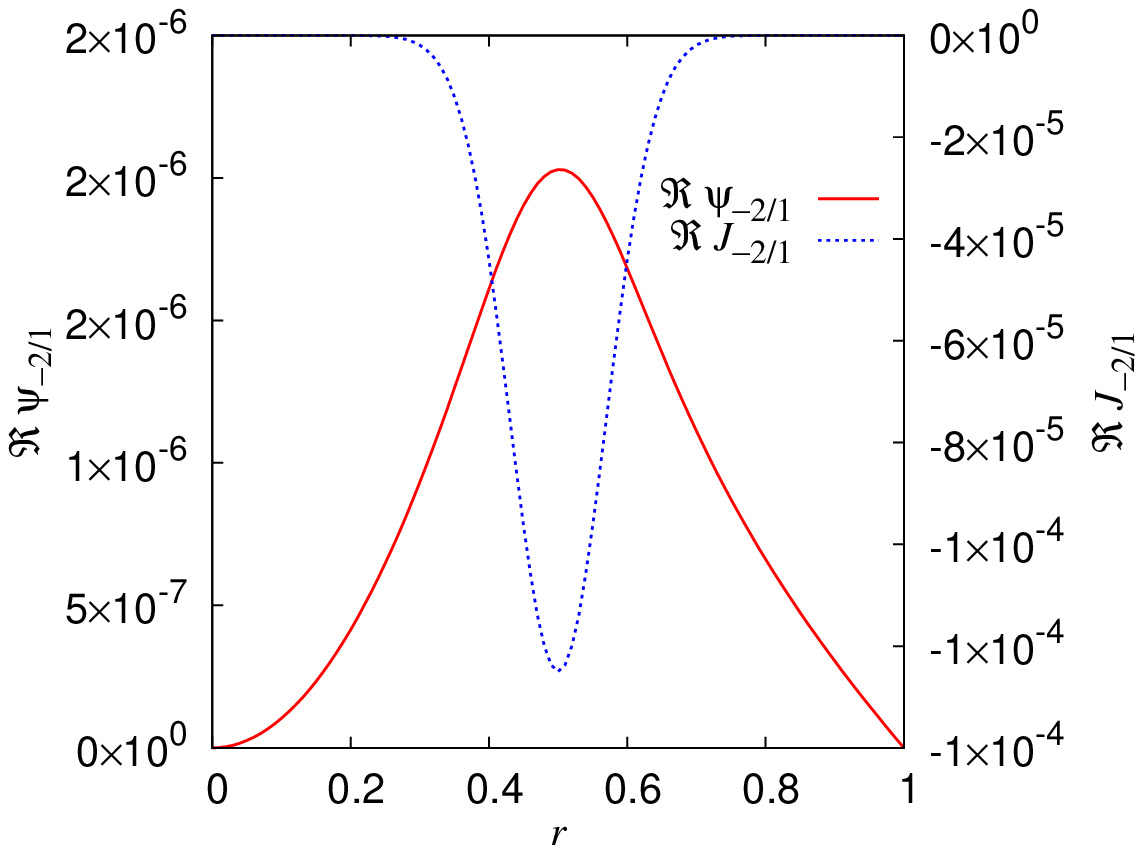}
   \label{fig:r-psi_r-J_r-mm2-n1-t000000_0000}
 } 
 \caption{Depiction of $m=-2$ and $n=1$ components of (a) $\Im U$ and $\Im \vphi$, 
(b) $\Re \psi$ and $\Re J$ at $t=0$.  A radial magnetic field
 exists at the $q=2$ surface.}
 \label{fig:initial-perturbation}
\end{figure}

Let us compare our   three  smoothing  kernels.  For the initial condition shown in
figures~\ref{fig:equilibrium} and \ref{fig:initial-perturbation}, the
right-hand sides of the evolution equations are plotted in
figure~\ref{fig:effect-kernel}. 
For SA-2 and SA-3, the amplitudes of the plotted figures are multiplied
by $10$ and $100$, respectively for easier comparison.  
Observe in figure \ref{fig:r-dUdt_i-mm2-n1-t000000_0000} how  the $r$-dependence is smoother for SA-3.

\begin{figure}[h]
 \centering
 \subfigure[]{
   \includegraphics[width=0.47\textwidth]{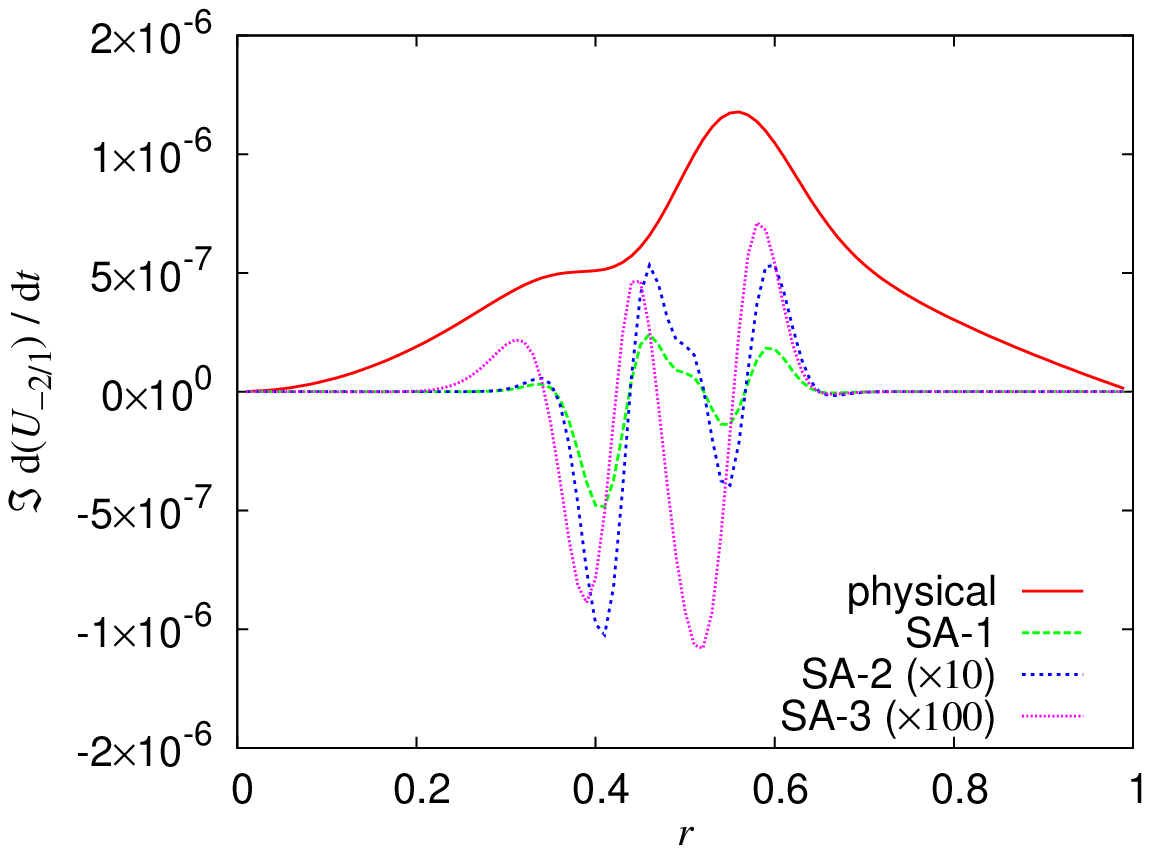}
   \label{fig:r-dUdt_i-mm2-n1-t000000_0000}
 } 
 \subfigure[]{
   \includegraphics[width=0.47\textwidth]{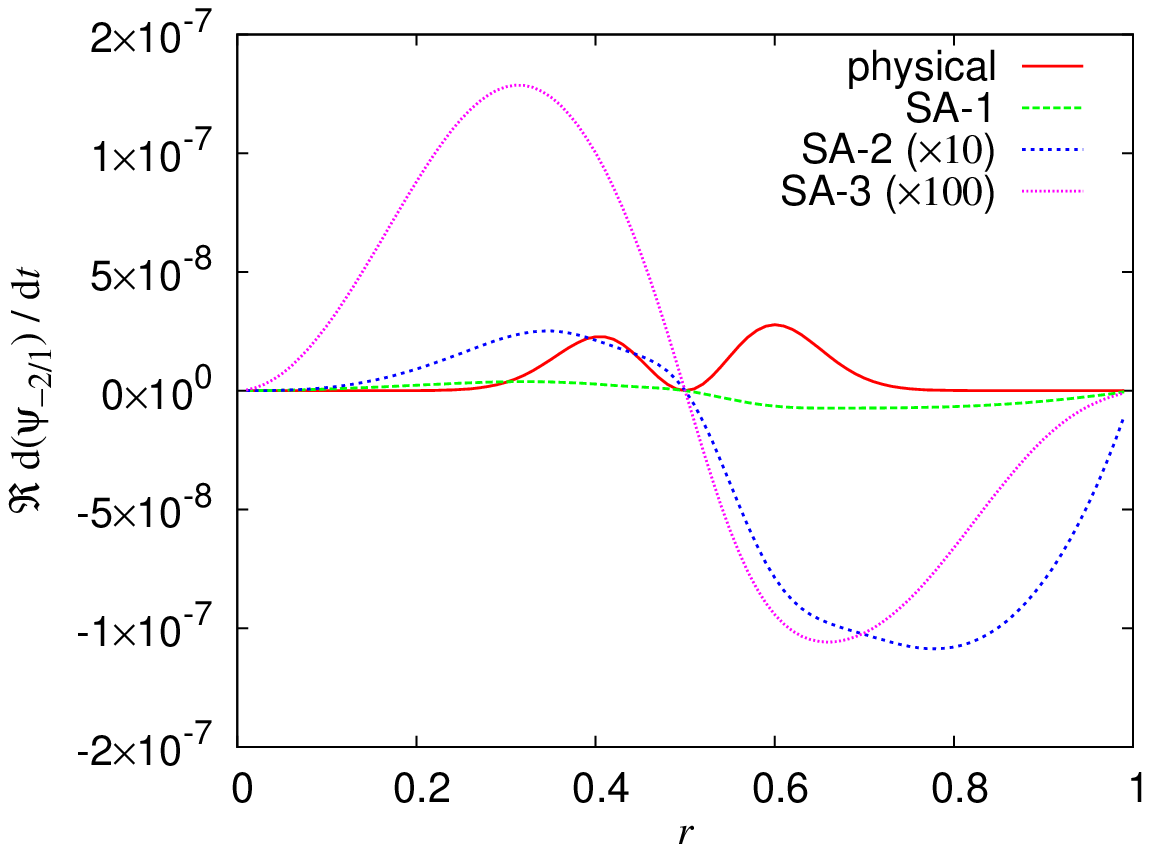}
   \label{fig:r-dpsidt_r-mm2-n1-t000000_0000}
 } 
 \caption{Depiction of $m=-2$ and $n=1$ components of (a) $\Im \frac{\td
 U_{-2/1}}{\td t}$ and (b) $\Re \frac{\td \psi_{-2/1}}{\td t}$ at $t=0$
 for physical dynamics and SA.  Since the amplitudes are
 largely different, those of SA-2 and SA-3 are multiplied by $10$ and
 $100$, respectively.  The significant smoothing effect in $r$ of SA-3 is observed  in (a).}
 \label{fig:effect-kernel}
\end{figure}

Figure~\ref{fig:r-g} shows the radial profile of $g_{m/n,-m/-n}(r,
r^{\pr\pr})$ of SA-3 for $r^{\pr\pr} = 0.2$, $0.4$, $0.6$ and $0.8$.
The mode numbers are $m=-2$ and $n=1$ in figure~\ref{fig:r-g-mm2-n1},
and $m=-10$ and $n=5$ in figure~\ref{fig:r-g-mm10-n5}.  
The range of the vertical axis is the same for both figures.  Note 
that $g_{m/n,-m/-n}(r, r^{\pr\pr})$ has smaller amplitudes for high $m$
and $n$.  This smoothing effect in 2D is the same  as that observed  in
\cite{Flierl-Morrison-11, Chikasue-15-PoP, Chikasue-15-JFM}, 
and is present also in SA-2.  As for the smoothing effect in $r$,
it is larger for smaller $m$ and $n$ because the radial extent of 
$g_{m/n,-m/-n}(r, r^{\pr\pr})$ is larger for smaller $m$ and $n$.  
Note that there is no smoothing effect in $r$ 
if $g_{m/n,-m/-n}(r, r^{\pr\pr}) = \delta (r - r^{\pr\pr})$, 
as in SA-2.  Thus the smoothing effect in $r$ of SA-3 may disappear
if $m$ and $n$ go to  infinity, while the smoothing  in $\theta$ and
$\zeta$ becomes infinitely large.

\begin{figure}[h]
 \centering
 \subfigure[]{
   \includegraphics[width=0.47\textwidth]{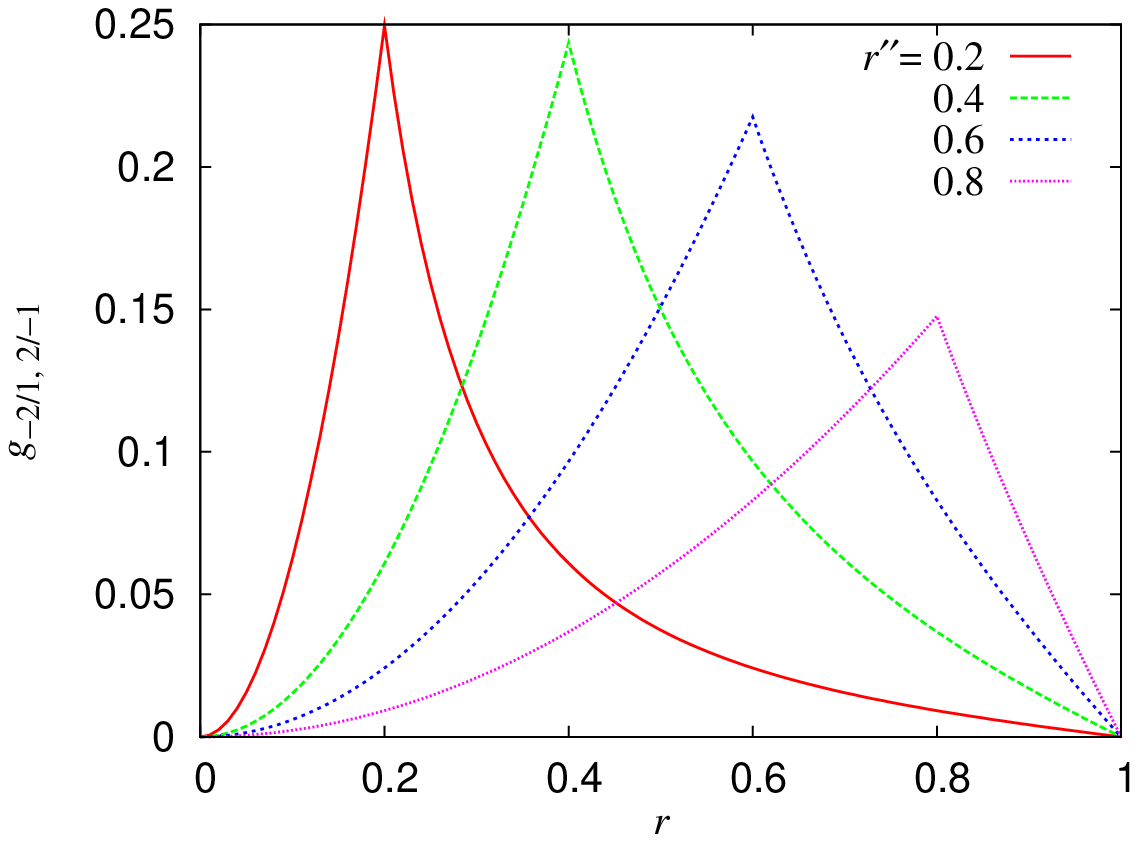}
   \label{fig:r-g-mm2-n1}
 } 
 \subfigure[]{
   \includegraphics[width=0.47\textwidth]{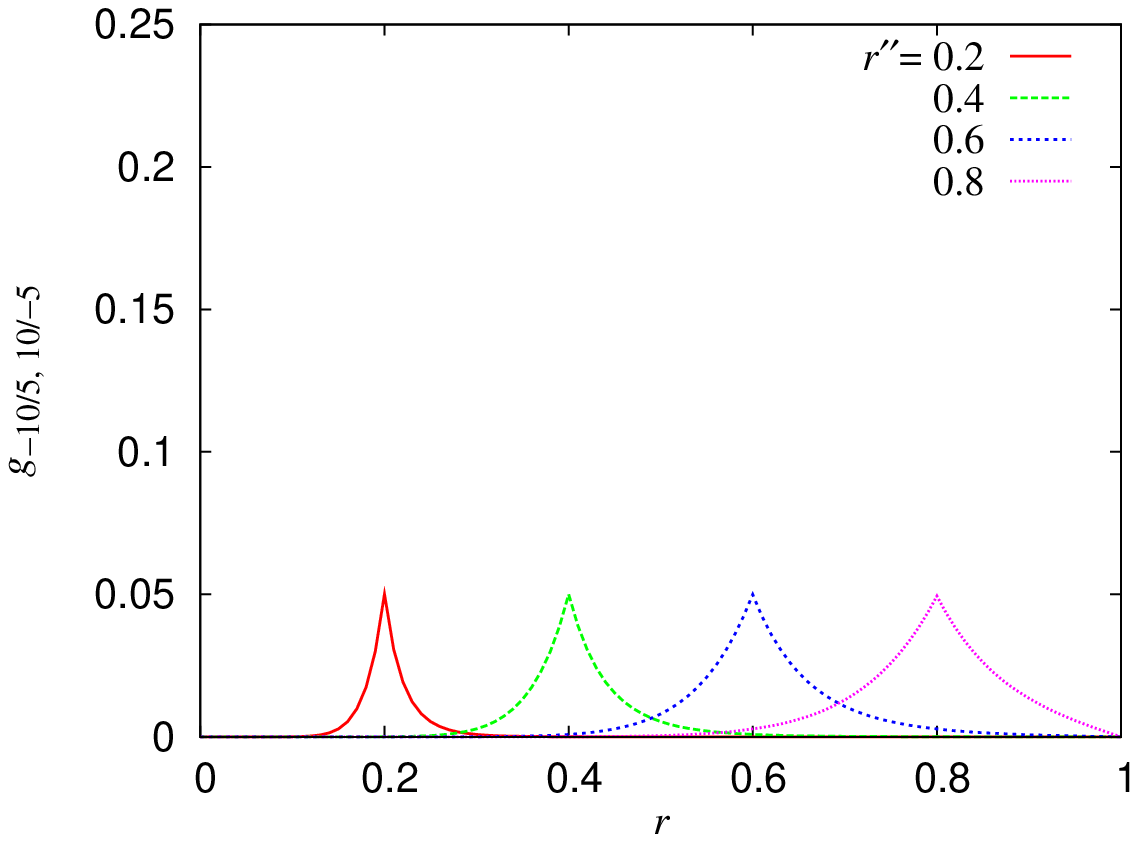}
   \label{fig:r-g-mm10-n5}
 } 
 \caption{Radial profile plots  of the 3D Green's functions 
$g_{m/n,-m/-n}(r, r^{\pr\pr})$ of SA-3 with 
(a) $m=-2$, $n=1$ and 
(b) $m=-10$ and $n=5$, for  $r^{\pr\pr} = 0.2$, $0.4$, $0.6$ and $0.8$.  
The amplitudes of $g_{m/n,-m/-n}(r, r^{\pr\pr})$ are smaller for larger
 $m$ and $n$, implying a  larger smoothing effect.  
Also, observe the larger smoothing  in $r$  for smaller $m$ and $n$, since  the
 radial extent of  $g_{m/n,-m/-n}(r, r^{\pr\pr})$ is larger for smaller $m$ and $n$.}
 \label{fig:r-g}
\end{figure}

The time evolution of the energy and the conserved quantities are 
shown in figure~\ref{fig:time-history} with 
$\alpha_{UU} = \alpha_{\psi\psi} = 100$   for SA-2 and SA-3.  
SA-1 was numerically unstable, and a stationary state was not obtained.
From figure \ref{fig:t-E}, we observe that the total energy 
$E_{\rm k} + E_{\rm m}$ decreases monotonically. 
Figure \ref{fig:t-dEdtoE} shows the time history of 
$\dis{
{| \td E_{\rm k} / \td t |}/{E_{\rm k}}
}$
and 
$\dis{
{| \td E_{\rm m} / \td t |}/{E_{\rm m}}
}$.
When they became lower than the tolerance $10^{-6}$, the simulation was
stopped.  Since the magnetic energy $E_{\rm m}$ is
dominant, its change is relatively small from the beginning. 
From figures \ref{fig:t-dCmoCm}, \ref{fig:t-Cv} and
$\ref{fig:t-Cc}$, we observe that quantities that should be conserved
are well conserved in the simulation.  
The change of $C_{\rm m}$ is monitored relatively to its initial value
$C_{\rm m}(0) = 3.76$, while $C_{\rm v}$ and $C_{\rm c}$ are  plotted directly since their initial values are zero.

\begin{figure}[h]
 \centering
 \subfigure[]{
   \includegraphics[width=0.47\textwidth]{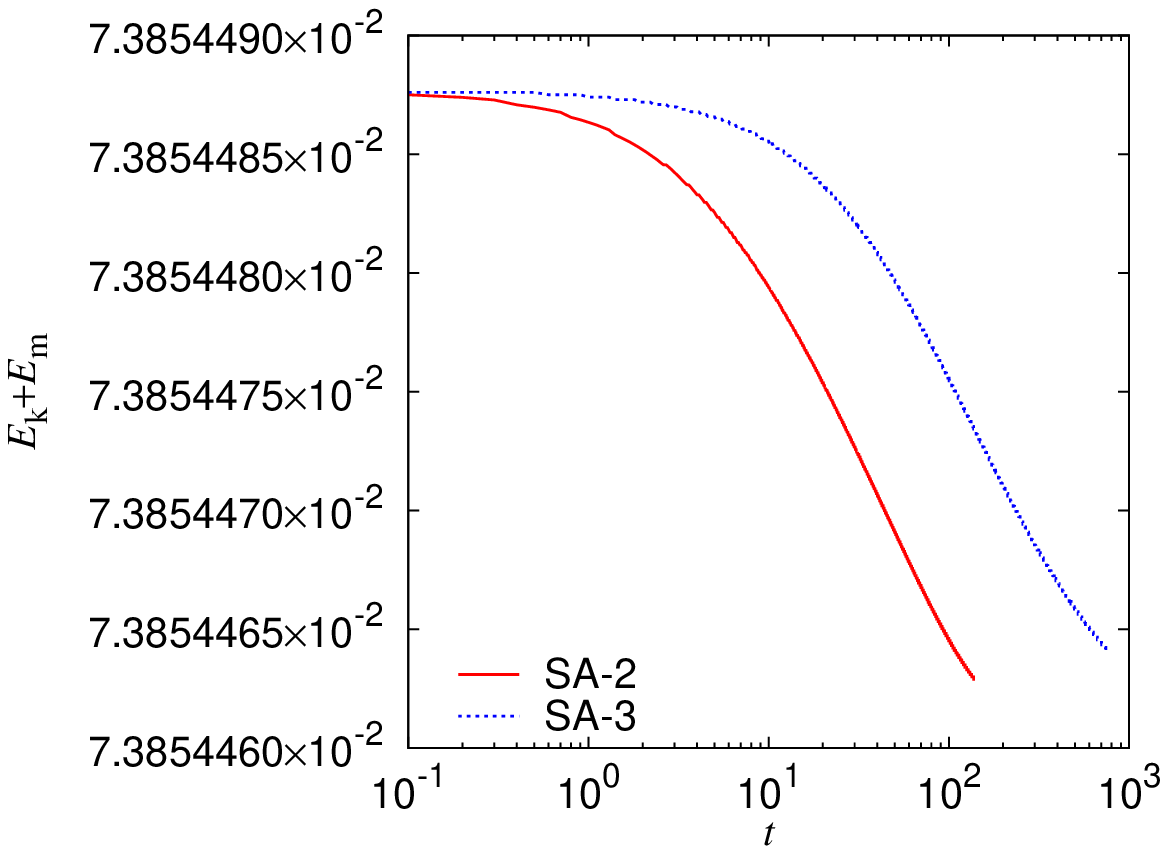}
   \label{fig:t-E}
 } 
 \subfigure[]{
   \includegraphics[width=0.47\textwidth]{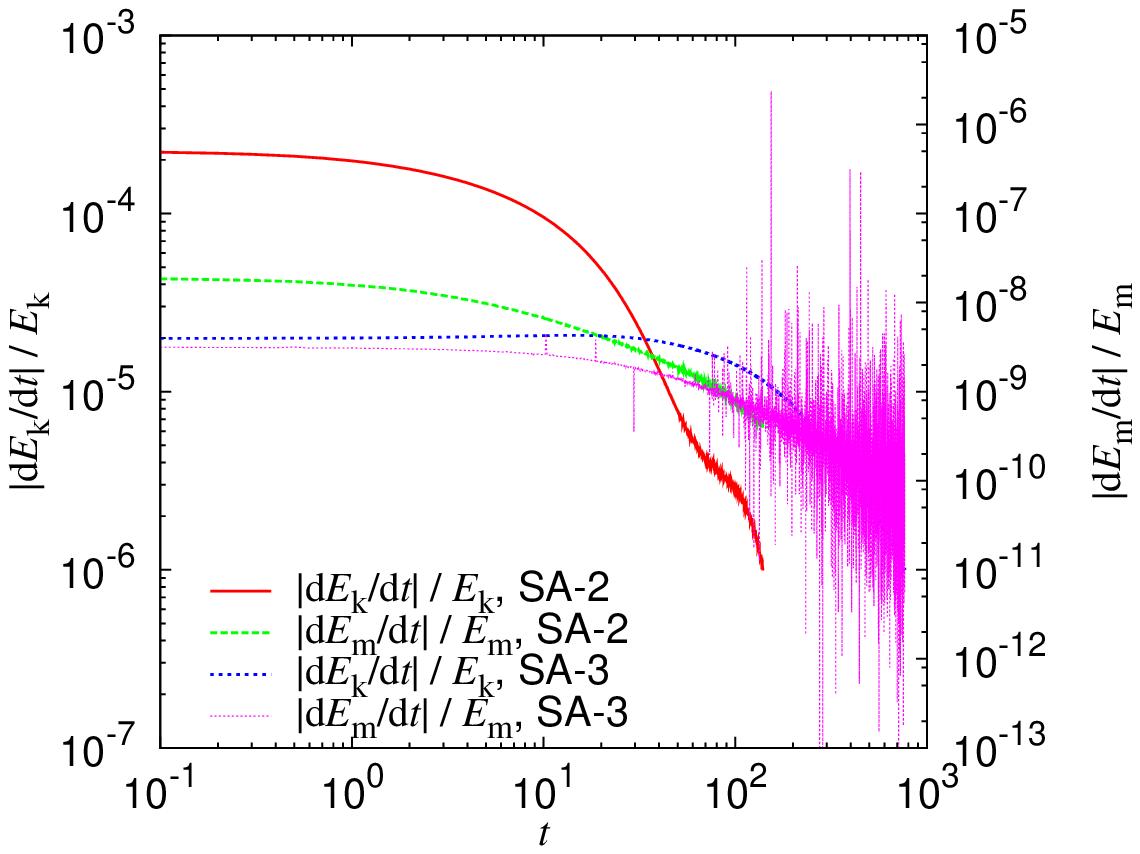}
   \label{fig:t-dEdtoE}
 } 
 \subfigure[]{
   \includegraphics[width=0.47\textwidth]{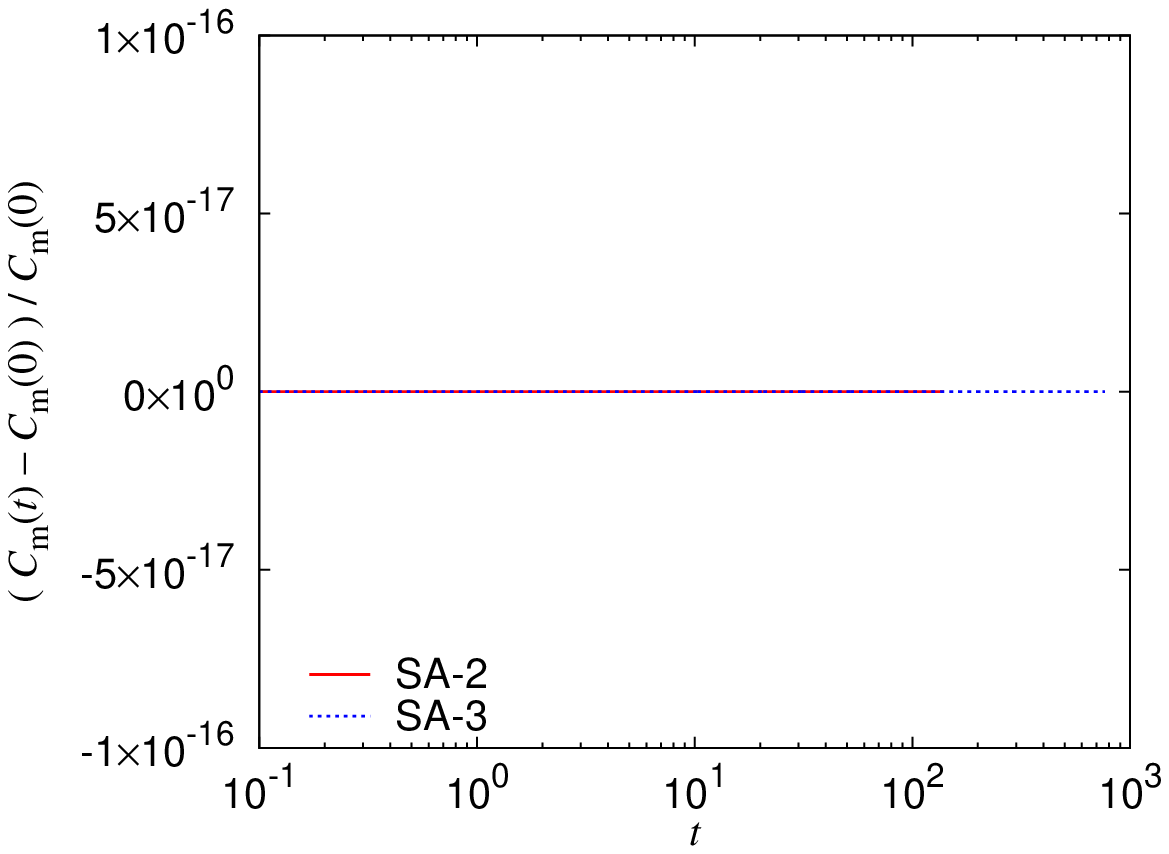}
   \label{fig:t-dCmoCm}
 } 
 \subfigure[]{
   \includegraphics[width=0.47\textwidth]{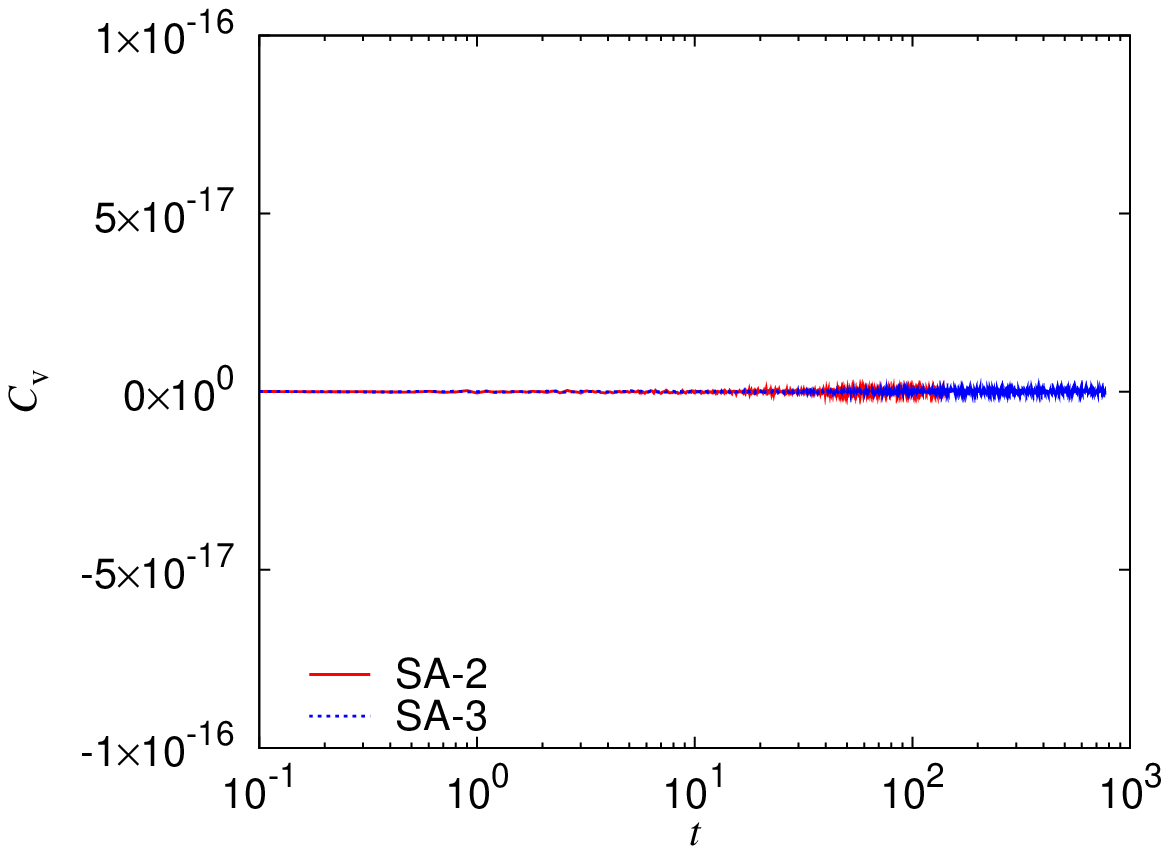}
   \label{fig:t-Cv}
 } 
 \subfigure[]{
   \includegraphics[width=0.47\textwidth]{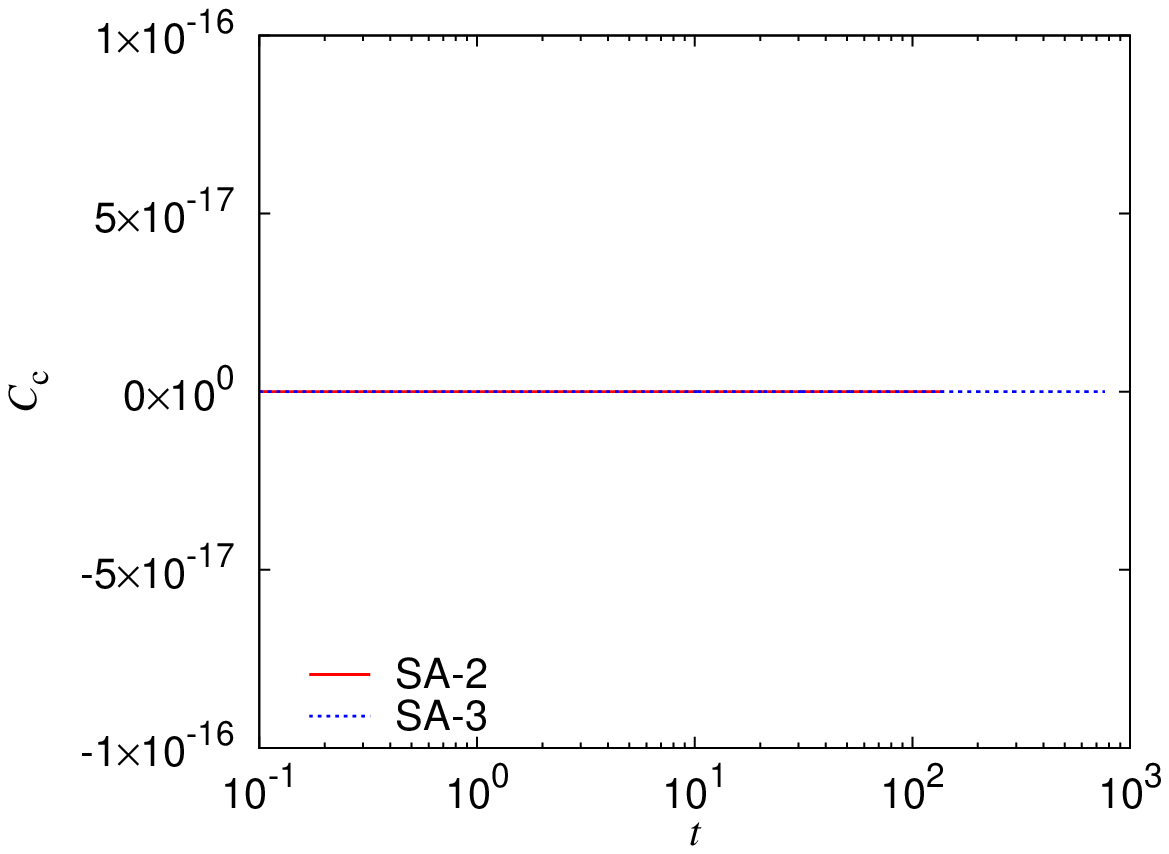}
   \label{fig:t-Cc}
 } 
 \caption{Time evolution of (a) total energy $E_{\rm k} + E_{\rm m}$,
 (b) relative change rate of energy $|\td E_{\rm k} / \td t| / E_{\rm k}$
 and $|\td E_{\rm m} / \td t| / E_{\rm m}$, 
 (c) relative change $( C_{\rm m}(t) - C_{\rm m}(0) ) / C_{\rm m}(0)$, 
(d) $C_{\rm v}$ and (e) $C_{\rm c}$.   The values 
$\alpha_{UU} = \alpha_{\psi\psi} = 100$ were used.  
The total energy decreases monotonically and reaches a stationary state.
The relative change of $C_{\rm m}$ is normalized by
 the initial value $C_{\rm m}(0)$ in (c).  Since $C_{\rm v} = 0$ and $C_{\rm
 c} = 0$ at $t=0$, just their values themselves are plotted in (d) and (e).}
 \label{fig:time-history}
\end{figure}

SA-3 requires  longer $t$ for   convergence.  Although the  time $t$ is not
physical and  depends  on $\alpha_{UU}$ and
$\alpha_{\psi\psi}$,   late convergence can also be because SA-3 smooths 
 in $r$, and thus  tends  to prevent generation of  fine structure in $r$.  Magnetic islands have
current channels that  may be easier to generate with  SA-2 than
SA-3.  This also indicates that a stationary state with fine structure
in $\theta$ and $\zeta$ may take more  simulation time 
using SA-2 and SA-3. 

Figure~\ref{fig:stationary-state-island} shows the real parts of the radial profiles 
$\Re \psi_{-2/1}$, $\Re \psi_{-4/2}$,  
$\Re J_{-2/1}$ and $\Re J_{-4/2}$ of the obtained stationary state.
The radial magnetic field of the  $m=-2$ and $n=1$ mode remains at the $q=2$
surface when the magnetic island exists.  These profiles differ greatly from the initial condition,
with larger amplitudes, while  the radial profile of the $m/n = -2/1$ mode is still similar to the
corresponding linear mode.   Therefore, the magnetic island of  this stationary state saturated in a
weakly nonlinear sense.

On the other hand, the radial profiles of SA-2  and SA-3 are a bit different.  
Also, almost no change was observed in $\Im U$ and $\Im \vphi$.
These will be discussed in the next section.

\begin{figure}[h]
 \centering
 \subfigure[]{
   \includegraphics[width=0.47\textwidth]{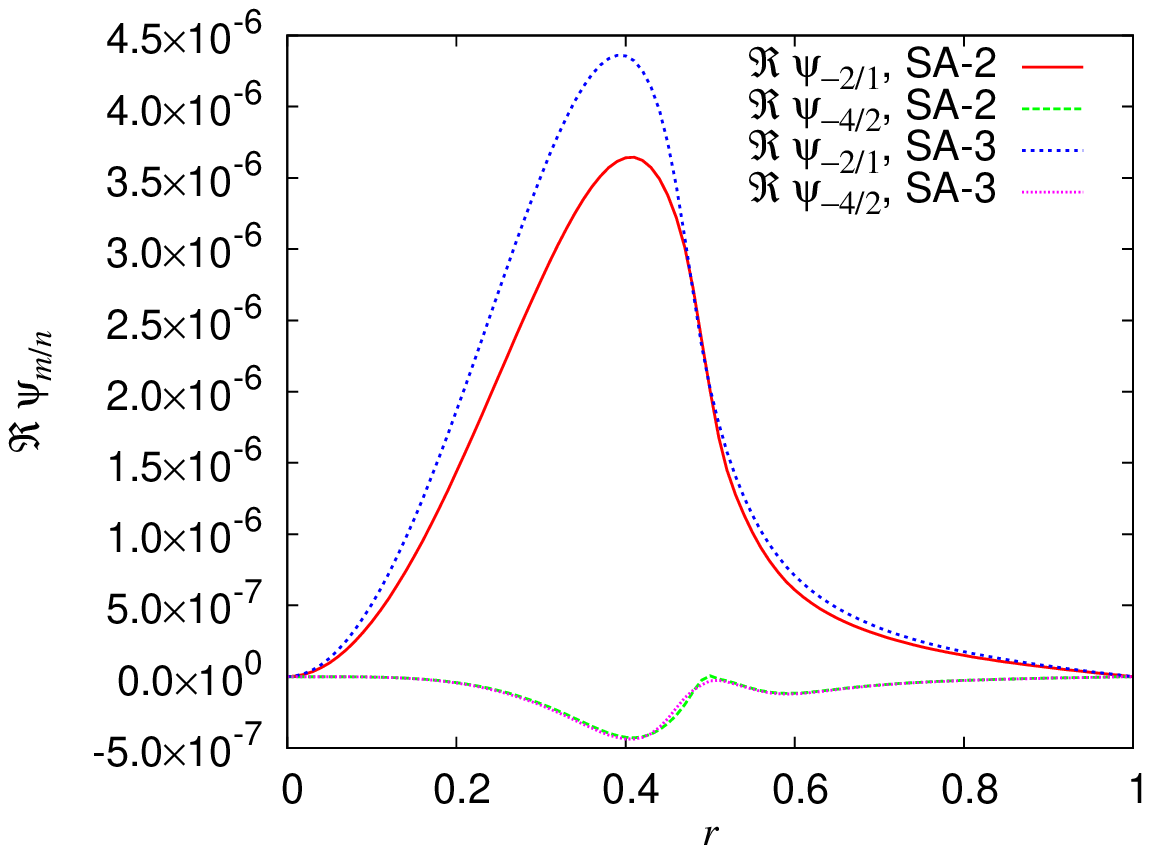}
   \label{fig:r-psi_r-final}
 } 
 \subfigure[]{
   \includegraphics[width=0.47\textwidth]{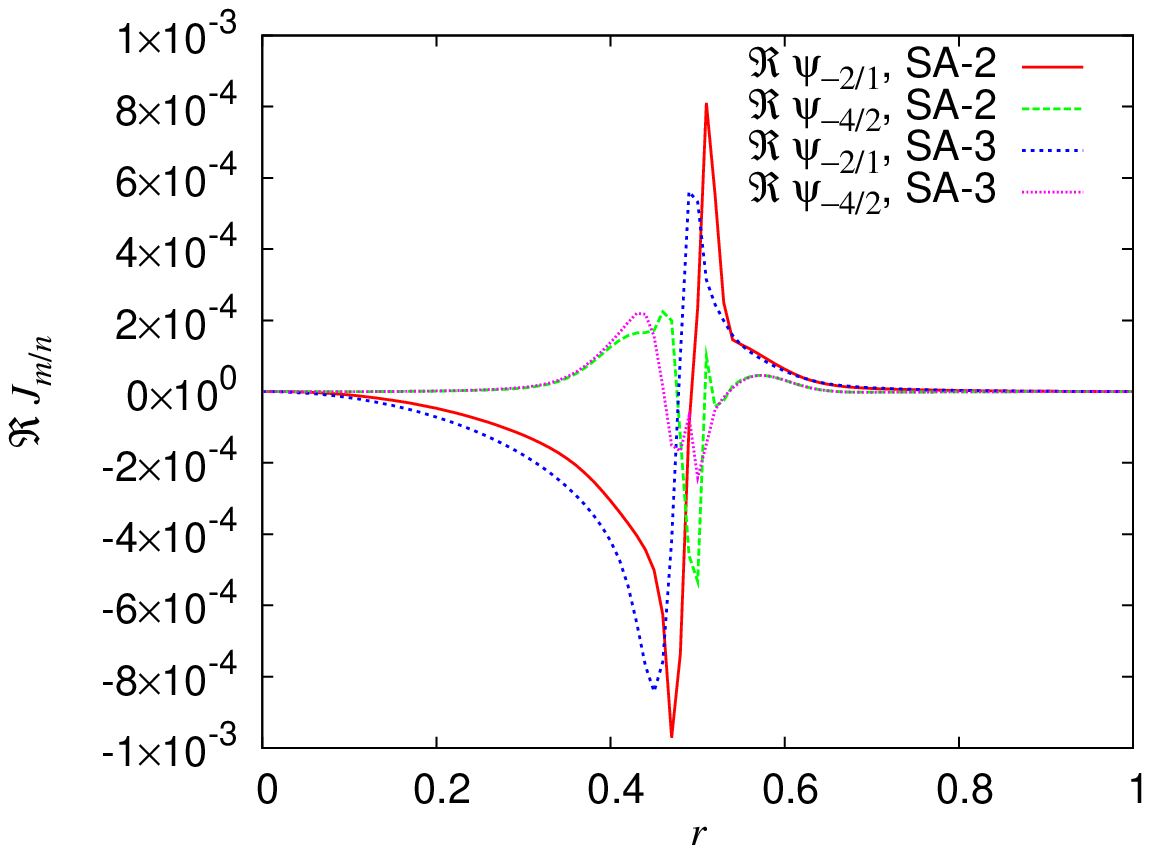}
   \label{fig:r-J_r-final}
 } 
 \caption{Radial profiles of 
(a) $\Re \psi_{-2/1}$ and $\Re \psi_{-4/2}$ and 
(b) $\Re J_{-2/1}$ and $\Re J_{-4/2}$ of the obtained stationary state are
 plotted.  Almost no changes occur  in $\Im U$ and $\Im \vphi$.  The $m/n=-2/1$
 components have similar structure as the linear mode.} 
 \label{fig:stationary-state-island}
\end{figure}

\section{Discussion}
\label{sec:discussion}

Firstly, let us investigate  why  SA-2 and SA-3 differ.  One reason could be  the tolerance for stopping the
simulation, which was set to 
$\dis{
| \td E_{\rm k} / \td t | / E_{\rm k}
}$
and
$\dis{
| \td E_{\rm m} / \td t | / E_{\rm m}
}$
becoming  smaller than $10^{-6}$.  While  the magnetic energy of the
$m/n=0/0$ component is very large,   the relative  rate of change of $E_{\rm m}$
 of the $m/n \neq 0/0$ components was very  small.  Therefore, we may
need another criterion for  convergence.  For example, separating out 
 the energy of the $m/n=0/0$ mode and monitoring the relative change rates 
 of both components of energy could be an improvement.  

Secondly, let us investigate  why $\Im U$ and $\Im \vphi$ did not
change during the SA evolution; i.e.,  why $\psi$ relaxed faster than $U$.
One possible reason is again the convergence
criterion.  If a longer simulation is performed with a much smaller
tolerance, $U$ and $\vphi$ may also change.  Another possible reason 
may be due to the choice of $\alpha_{UU}$ and $\alpha_{\psi\psi}$,
especially their ratio.  
If we write the evolution equations for SA-1 explicitly, we have
\begin{eqnarray}
 \frac{\pd U}{\pd t}
=
 \alpha_{UU} [ U(\vx), f^{U}(\vx) ] 
 + \alpha_{\psi\psi} \left(
     [ \psi(\vx), f^{\psi}(\vx) ] 
     - \veps \frac{\pd f^{\psi}(\vx)}{\pd \zeta}
                     \right),
\\
 \frac{\pd \psi}{\pd t}
=
 \alpha_{\psi\psi} \left(
     [ \psi(\vx), f^{U}(\vx) ] 
     - \veps \frac{\pd f^{U}(\vx)}{\pd \zeta}
                     \right). 
\end{eqnarray}
Therefore, the ratio of $\alpha_{UU}$ to $\alpha_{\psi\psi}$ can
significantly affect the time evolution of $U$.  
As was studied in \cite{Chikasue-15-PoP} for the 2D cases, 
the relaxation path can change
if we change the ratio of $\alpha_{UU}$ to $\alpha_{\psi\psi}$.  As we
observe, the time evolution of $U$ is governed by two advection fields
$f^{U}$ and $f^{\psi}$, while $\psi$ by $f^{U}$ only.  Therefore the
relaxation path can change if we change the ratio of contributions from
$f^{U}$ and $f^{\psi}$.  
This situation is also the same for SA-2 and SA-3.
If the relaxation of $U$ is much slower than $\psi$, a simple  solution
is to increase the ratio of $\alpha_{UU}$ to $\alpha_{\psi\psi}$.  Then
the right-hand side of the evolution equation of $\psi$ becomes smaller
and that of $U$ becomes larger.  
However, if $\alpha_{UU}$ is increased in the present code, the
simulation tends to be unstable. 
The dependence of the numerical stability on $\alpha_{UU}$ and
$\alpha_{\psi\psi}$,  in addition to the choice of the symmetric bracket, 
needs to  be examined more carefully. 

The result of  section \ref{sec:result} is only one example of  a magnetic island 
stationary state achievable with SA.  When the
initial perturbation of  the $m=-2$ and $n=1$ component was chosen larger, the $m=0$ and $n=0$ components of $U$ and $\psi$ were changed significantly by the nonlinear effects, leading to a different
stationary state.  Incorporating Dirac constraints as in \cite{Flierl-Morrison-11} should be explored in the future for  selecting out desired states.   
Also, the  effects of the $m=0$ and $n=0$ 
component of the plasma rotation should be investigated because it
changes the linear stability against tearing modes.  Details of
these issues will be studied and will be reported on  in the near future.

\section{Summary}
\label{sec:summary}

The method of simulated annealing (SA) was developed to obtain a
three-dimensional stationary state of low-beta reduced MHD in
cylindrical geometry.   The theory of  SA was explained for  low-beta reduced MHD, and
three versions of the symmetric bracket were introduced.
A simulation demonstrated that the energy of the system monotonically
decreases  by  SA, while  conserving other invariants.    Starting from a cylindrically  symmetric state with the  addition of a  perturbation that opens a small magnetic island at a rational surface, SA generated a stationary state with magnetic islands as a lower energy state.  
Smoothing effects by the symmetric brackets were also examined.  
A symmetric bracket with higher smoothing  may require longer 
simulation time for convergence, while it can contribute to  numerical
stability.   Several issues for  consideration in the future  were discussed.

\ack
MF   was supported by JSPS KAKENHI Grant \#23760805 and \#15K06647.
PJM was supported by U.S.\  DOE Grant   \#DE-FG02-04ER-54742 and  the Humboldt Foundation.

\section*{References}
\bibliographystyle{prsty}

\end{document}